\documentclass{ws-procs975x65}

\usepackage{amsmath}
\usepackage{amsfonts}
\usepackage{amssymb}
\usepackage{dsfont}
\usepackage{latexsym}
\usepackage{epsfig}

\def\beq{\begin{equation}}
\def\eeq{\end{equation}}
\def\munu{{\mu\nu}}

\def\bitem{\begin{itemize}}
\def\eitem{\end{itemize}}
\def\d{\partial}

\def\5M{\mathcal{M}^5}

\def\2f{{\phi}^2}
\def\teta{\vartheta}
\def\i{\hat{I}}
\def\j{\hat{J}}

\def\ds5{d s_{5}}
\def\bear{\begin{array}}
\def\ear{\end{array}}

\begin{document}

\title{Hamiltonian formulation of the 5-D Kaluza-Klein model and test-particle motion}

\author{Valentino Lacquaniti$^{*\diamond}$ and Giovanni Montani$^{*\dag}$}

\address{$^*$ICRA---International Center for Relativistic Astrophysics\\
Dipartimento di Fisica (G9), Universit\`a  di Roma, ``La Sapienza",\\
Piazzale Aldo Moro 5, 00185 Rome, Italy.}
\address{$^{\diamond}$ Dipartimento di Fisica "E.Amaldi", Universit\'a di Roma "RomaTre", \\ Via della Vasca Navale 84, 00146 Rome, Italy.}
\address{$^\dag$ENEA-C.R. Frascati (U.T.S. Fusione),\\ via Enrico Fermi 45, 00044 Frascati, Rome, Italy.}

\address{E-mail: lacquaniti@fis.uniroma3.it, valentino.lacquaniti@icra.it \\ montani@icra.it}

\begin{abstract}
We examine the ADM reformulation of the 5-D KK model: the dimensional reduction is provided to commute with the ADM splitting and we show  how the time component of the gauge vector is given by combination of the Lagrangian multipliers for the 5-D gravitational field. We consider 5D particles motion and after dimensional reduction the definition of charge is recovered within electrodynamic coupling. A  time-varying fine structure constant is recognized because an extra scalar field is present in the 4-D theory.\end{abstract}
\bodymatter

\section*{}
KK 5-D model provides a unified geometrical picture of gravitation with electromagnetism plus an extra scalar field $\phi$ . This picture is achieved via the assumption of some restrictives hyphotesis on the 5-D space-time that break the 5-D Poincar\'e symmetry \cite{kaluza} \cite{klein1},\cite{acf},. So a first goal is to understand if the KK reduction is compatible with AMD splitting, that is usually requested in canonical approach, in order to compute the Hamiltonian of the model \cite{adm}. Due to the symmetry breaking we could have not equivalent dynamics wheter or not the ADM splitting is performed before we implement the KK reduction. We have two possible procedures (we label these as KK-ADM and ADM-KK ).
In KK-ADM procedure we perform as first the usual KK reduction of the metrics, and then a $3+1$ ADM splitting of the gravitational tensor and the  abelian gauge vector:
$$
J_{AB}  \Rightarrow \left\{
\bear{l}
g_{\munu} \rightarrow  \teta_{ij},S_i,N \\
A_{\mu}  \rightarrow  A_i,A_0 \\
\phi \rightarrow   \phi \\
\ear
\right.
\rightarrow
\left(
\begin{array}{ccc}
 N^2-S_iS^i-\2f A_0^2   &\quad  -S_i-\2f A_0A_i      & \quad -\2f A_0 \\
 -S_i^2\2f A_0A_i         &\quad -\teta_{ij}-\2fA_iA_j  &\quad  -\2f A_i \\
-\2f A_0                    & \quad -\2f A_i                   &\quad  -\2f
\end{array}
\right)
$$
Here $J_{AB}$ is the 5D metrics, $A_{\mu}$ the gauge vector, $N,S_i,\teta_{ij}$ the \textit{Lapse} function, the 3D \textit{Shift} vector and the 3D induced metrics ($A,B =0,1,2,3,5;\quad \mu,\nu= 0,1,2,3;\quad i,j=1,2,3$).
In this way we have a not complete space-time slicing, due to the fact that we are doing a $3+1$ splitting in a 5-D ambient, so the extra dimension is not included . In the ADM-KK procedure we deal as first with a $4+1$ splitting that includes the extra dimension and after consider the KK reduction related to the pure spatial manifold:
$$
J_{AB}  \Rightarrow  \left\{
\bear{l}
h_{\i,\j} \, \rightarrow  A_i,\teta_{ij},\phi \\
N_{\i} \,\,\, \rightarrow  N_i, N_5 \\
N \,\,\,\, \rightarrow  N 
\ear
\right.
\Rightarrow
\left(
\begin{array}{ccc}
N^2-h_{\i\j}N^{\i}N^{\j} &\quad -N_{i} &\quad -N_5 \\
-N_{i} &\quad -\teta_{ij}-^2{}\2f  A_iA_j &\quad\ -\2f A_i \\
-N_5 &\quad -\2f A_{i} &\quad -\2f
\end{array}
\right)
$$
Here $N_{\i}$ and $h_{\i\j}$ are the 4D \textit{Shift} vector and the 4D spatial induced metrics ($\i,\j=1,2,3,5$).
Now we have a complete slicing but the set of variables lacks the component $A_0$. Hence, both of procedures are unsatisfactory and we don't know if they commute. Despite the outcomes of the metrics seem to be differents , nevertheless we are dealing  with objects that must show well defined properties under pure spatial KK diffeomorphism. This allow us to look for "conversion formulas" between our two metrics. Indeed we can implement the KK reduction on $N_{\i}$; by this we recognize that $N_{\i}$ is not a pure 4D spatial vector neither simple gauge vector but a mixture of them. A detailed study of the tetradic structure allow us to state the following formulas for $N_{\i}$ which map metrics into each other.
$$
\left\{
\begin{array}{l}
N_{i}=S_{i}+ \phi^2A_0A_{i} \\
N_5=\2fA_0
\end{array}
\right.
\quad\quad\quad
\left\{
\begin{array}{l}
N^{i}=S^{i} \\
N^5=N^2A^0
\end{array}
\right.
$$
The real physical meaning of these formulas is provided once we compare the two Lagrangian. By our formulas we can recast the lagrangians in the same set of variables and we get:
$$
L_{adm-kk}=L_0 +\frac{1}{8\pi G} \sqrt{\teta}NK\d_{\eta}\phi \quad\quad
L_{kk-adm}=L_0+\frac{1}{8\pi G} \sqrt{\teta}N\d_{\eta}\d_{\eta}\phi
$$
Here $K$ is the trace of the 3D extrinsic curvature and we have $\d_{\eta}=(1/N)(\d_0-S^i\d_i)$.
$L_0$ is a term, egual for both Lagrangians, that contains no derivatives of $\phi$; the remaining contributes are found to be equivalent apart from surface term. Hence, by assuming proper boundary conditions, in order to discard these terms, we conclude that we are dealing with equivalent dynamics and that our conversion formulas have a real physical meaning. Therefore we can conclude that ADM splitting commutes with KK reduction; we are able in both cases to have a complete space time slicing and to rebuild the time component of the gauge vector. Moreover, due to the commutation, we can have a unique, well-defined, Hamiltonian $H$; it can be proved that $H$ can be casted in an explicit form like the following one: $H=NH^N+S_iH^i+A_0H^0$.
Here $H^N$ is the \textit{Superhamiltonian}, $H^i$ the \textit{Supermomentum}, $H^0$ the electrodynamical contribute
; $A_0$ appears like a lagrangian multiplier and it is  predicted by the conversion formula $A_0\infty N_5$; this means that after the KK reduction one of the geometrical constraints of the model becomes a gauge one.
So the proof of commutation is a positive check for the consistency of the KK model, is necessary in order to consider the hamiltonian formulation and provides  insight in the  understanding of the gauge simmetry generation. Finally, it can be a first step in the Ashtekar reformulation of KK model. For an accurate discussion about these results see ref. \cite{lm}.
Another point of interest is the presence of $\phi$ in the model and its role in the dynamics. Indeeed there are some reasons to think that $\phi$ can be a good time variable in the relational point of view and hence solve the problem of the frozen formalism. An insight on the role of $\phi$ is provided by the study of test-particles motion.
By studying the 5D geodesic equations  we are able to reproduce the usual electrodynamics for a test-particle in a 4D space-time, where the charge-mass ratio is ruled out as follows: $q/m=u_5(1+\frac{u_5^2}{\phi^2})^{-1}$.
In this formula $u_5$ is the fifth covariant component of the 5D velocity and can be proved that it is a constant of motion and a scalar under KK transformations. In general, the charge is not conserved due to the presence of $\phi$, and at the same time we cannot be sure that our test particle has a realistic $q/m$. A large scalar field ($\phi>10^{21}$ for the electron ) allow us to have realistic value for the charge mass ratio avoiding the problem of planckian mass, and,  moreover, allow us to restore the conservation of charge at a satisfactory degree of approximation. Actually, a time-varying charge is very interesting; an isotropic, slow varying $\phi$ can explain the time-variation of the fine structure constant over cosmological scale which seems to be inferred by recent analysis \cite{alfa}. Other features of $\phi$ arise from the study of  Friedmann/DeSitter-like models; we deal  only with $\phi(t)$ and the 3D scale factor $a(t)$ ; our results show that in both cases $\phi$ mimics the behaviour of the scale factor and can be increasing or decreasing in time with power law (Friedmann) or exponential law (DeSitter).
Hence, the scalar field shows some classical feature that we can link to a possible role of it as a time -like variable.
Seems interesting to deal with induced matter theory (and with a detailed cosmological solution ) and/or with Ashtekar reformulation of variables in order to deep understand the role of $\phi$.

\end{document}